# En route to a multi-model scheme for clinker comminution with chemical grinding aids

R. K. Mishra[1], D. Geissbuhler[2], H. A. Carmona[1,3], F. K. Wittel[1], M. L. Sawley[2], M. Weibel[4], E. Gallucci,[4] H. J. Herrmann[1], H. Heinz[5], R. J. Flatt[1*]

[1]Institute for Building Materials, ETH Zurich, 8093 Zurich, Switzerland
[2]ACCES, School of Engineering, EPFL, Station 11, 1015 Lausanne, Switzerland
[3]Departamento de Fisica, Universidade Federal do Ceara, 60451-970 Fortaleza, Brazil
[4]Sika Technology AG, Zurich, Switzerland
[5]Department of Polymer Engineering, University of Akron, OH-44325, Akron, USA
*flattr@ethz.ch

We present a multi-model simulation approach, targeted at understanding the behavior of comminution and the effect of grinding aids (GAs) in industrial cement mills. On the atomistic scale we use Molecular Dynamics (MD) simulations with validated force field models to quantify elastic and structural properties, cleavage energies as well as the organic interactions with mineral surfaces. Simulations based on the Discrete Element Method (DEM) are used to integrate the information gained from MD simulations into the clinker particle behavior at larger scales. Computed impact energy distributions from DEM mill simulations can serve as a link between large-scale industrial and laboratory sized mills. They also provide the required input for particle impact fragmentation models. Such a multi-scale, multi-model methodology paves the way for a structured approach to the design of chemical additives aimed at improving mill performance.

## Introduction

Comminution is a very energy intensive production step in the cement manufacturing, which involves the operation of large mills to achieve a desired clinker particle size distribution (PSD). In order to optimize this process, we focus on physical and chemical aspects of grinding. Industrial cement ball mills consist of a rotating dual-chamber cylinder sized up to 8 meters in diameter and 20 meters in length. Such mills are driven by electric motors in the range of 20 – 40 MW, consuming 30% of the overall electrical energy dedicated to the entire production chain of cement.[1-4] Most of the earlier work on modeling of clinker grinding was focused on the simulation of the charge motion using the Discrete Element Method. Until now numerical models based on different physical scales present in clinker and milling systems combining physicochemical aspects have not been possible to integrate into a single methodology. Various kinds of simulated systems described in this study contribute to construct a framework that captures clinker grinding in an unconventional way.

Comminution theory, as developed by process engineers, treats mills as "black boxes", where input parameters such as speed of rotation, loading, separator efficiency and others are factors that are combined in a way, such that the size reduction is related to the energy consumption.[5, 6]

Such approaches are used for the design of mills; however parameters are only interpolated within the range of experience.[7] What is missing is the physics and chemistry behind the milling process. The problem becomes particularly challenging by the interaction between processes across length scales of 10 orders of magnitude difference - from the atomistic scale, where surface formation and interactions rule, to the macroscopic scale of the machinery involving the grinding media kinematics. The mill capacity is a function of the mill, the liner design, the use of different grinding media, the charge and material grindability, as well as of organic grinding aids (GAs). Academic and industrial researchers alike have been faced with the problem of isolating phenomena occurring on a reduced scale that are both accessible to available analysis techniques and representative for the complete problem.

In principle the problem is too difficult to solve from a theoretical point of view in a closed form. Nevertheless we exemplify atomistic force field and (micro) mechanical models based on the Discrete Element Method on different scales and explain how they can be enriched by findings from models of smaller scales (see Fig. 1). In situ testing of grinding aids in industrial cement mills under stationary conditions can only be reached after many tons of clinker are milled, rendering screening studies both costly and impractical. As a consequence, such experimental studies are often conducted in laboratory-size tumbling mills (e.g. ball or planetary mills) containing grinding media that differ in size, shape and material. Since the ball kinematics may therefore not be comparable, it is unclear whether results obtained on these order-of-magnitude smaller mills still hold for the full-scale cement mill. Obtaining suitable correspondence is far from trivial, as high efficiency separators (HES) and different process temperatures change the problem significantly. Molecular modeling can give new insights, which are not available experimentally or theoretically but essential to understand working mechanisms of GAs for improved efficiency.

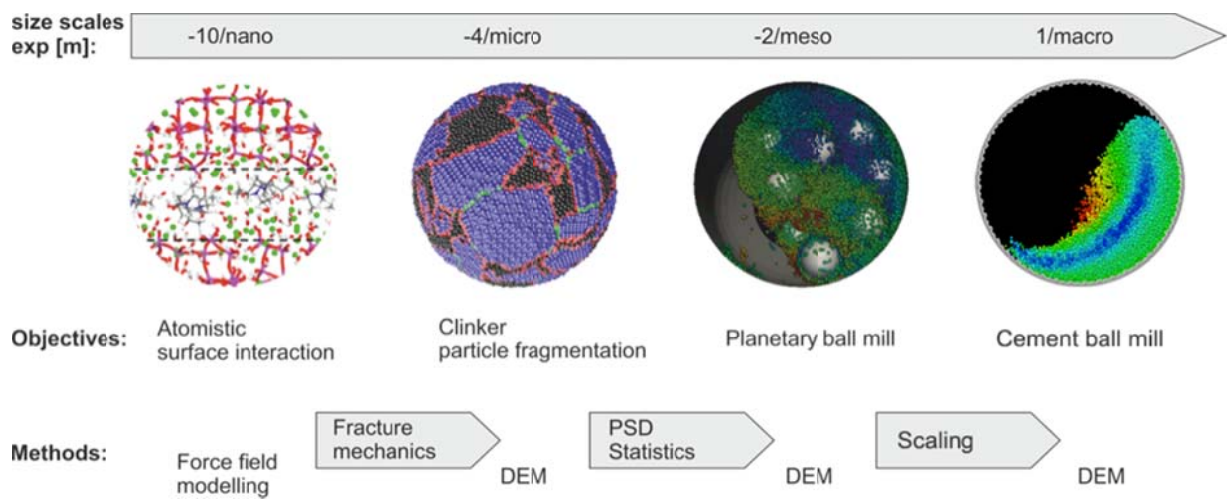

**1  Scales, objectives and methods involved in the multi-model scheme.**

The paper is organized as follows. First we describe simulations with GAs and clinker components in atomistic force field models. Subsequently, up-scaling of results over 6 orders of magnitude using fracture mechanics[8, 9] leads to an idealized clinker model for fragmentation simulations. This micro-scale model incorporates the clinker as a multi-crystal material and can be used to simulate the outcome of typical situations for which fragmentation in mills occurs. The computed kinematics of a (meso-scale) laboratory mill and full-scale industrial cement ball mill, expressed in terms of impact energy distribution (IED) of the particle charge, are then presented. Details are provided for further information required to be exchanged between different levels that should provide a suitable basis for a complete implementation of this multi-level multi-model approach. Finally, we summarize and give an outlook on possible future achievements in this field.

## Atomistic force field simulations of clinker with grinding aids

Design of efficient grinding aids includes many factors such as, conformations of molecules on clinker surfaces, combination of polar functional groups, hydrophilic/hydrophobic ratios and molecular size etc.[1, 10] In this paper, the main implications of molecular dynamic (MD) studies is to understand the interactions mechanisms of organic molecules with cement minerals. We are involved in the development and validation of INTERFACE force field parameters for both hydrated and unhydrated cement minerals that accurately reproduce density, lattice constants, cleavage energy, solid−liquid interfacial tension, as well as mechanical and structural properties with respect to experiment.[10-12] Note that, compared to alternative models[13-15], our approach captures the important surface polarity and interfacial properties correctly.

We focus on the four important phases of ordinary Portland cement (OPC) clinker, namely tricalcium silicate ($C_3S$) with Mg, Al, and Fe impurities in the low percent range (50−70%), dicalcium silicate ($C_2S$) containing similar metal substitutions (15−30%), tricalcium aluminate ($C_3A$) in weight percentage of 5−10, and tetracalcium aluminoferrite ($C_4AF$) in weight percentage of 10−15.[16] Until now, we have validated force field models of pure $C_3S$ (main phase of cement) and $C_3A$ (most-reactive phase).[10, 12] We already concluded that surface polarity of $C_2S$ is lower than $C_3S$ due to the absence of oxide ions and also due to more covalent character. We also found in a preliminary simulation study that the cleavage energy of $C_2S$ is lower than $C_3S$ phase.[17] Ferrite ($C_4AF$) is one of the most complex clinker phases in the cement and focus of current molecular dynamics studies. Computed properties of $C_3S$ and $C_3A$ phases are summarized in Table A1. The simulation methodology and force field validations are discussed in detail in our earlier papers on $C_3S$ and $C_3A$.[10, 12]

Chemical GAs significantly improve mill efficiency, characterized in terms of the specific surface area using Blaine fineness and sieve residues.[1] They primarily shift the particle size distribution curve towards smaller diameters but do not seem to influence the shape of the curve considerably. Because of optical and practical restrictions neither measurements in air nor in solvents delivers distribution curves beyond the scattering of laser particle size distribution (PSD) analysis. Using MD simulations, we relate adsorption and agglomeration energies to rank the GAs on the basis of interaction strengths with clinker surfaces as well as reduction in agglomeration energies. The agglomeration energy is the released energy when freshly cleaved surfaces (particles) come together in order to reduce specific surface area and thus cleavage/surface energy.[10] Grinding aids (mostly organic compounds) reduce the surface energy of the clinker and therefore also the energy that is released when particles agglomerate. The energy consumption decreases during grinding as a consequence of various physical and chemical phenomena happen simultaneously during the clinker milling in the presence of grinding aids.[1]

The chemical structure of selected commercial GAs (TIPA, TEA, MDIPA, and Glycerine) and other organic compounds (DPG, DPGMME, and DPGDME) is shown in Schemes A1 and A2. Models of hydroxylated (Hyd.) $C_3S$ and $C_3A$ surfaces are used for the simulations because the majority of commercial clinkers are mostly hydroxylated.[1, 10, 12] Interactions of DPG on hydroxylated $C_3S$ (HC) surface is exemplarily shown in Figure 2a. The hydroxyl groups of the hydroxylated $C_3S$ surfaces are close to the polar groups of organic molecules. Adsorption of GAs occurs on the ionic clinker surfaces mostly due to hydrogen bonds, complexation of superficial calcium ions and dipolar interactions. DPGMME, DPG, TEA, TIPA, MDIPA, and glycerine are adsorbed more due to hydrogen bond formation. For these molecules we find strong hydrogen bonds due to shorter O···H bond length (0.17 nm ± 0.03) and H—O···H bond angles close to 180 degree. Additionally a weak hydrogen bond with a longer distance (0.25 nm ± 0.05) length and bond angle far from 180 degree is found. The strength of calcium ion complexation with alcohol based GAs depends upon the coordination number or number of nearest calcium ions. The adsorption strength (in kJ/mol/molecule) was found in the order TEA (–75 ± 8) < TIPA (–100 ± 15) ~ DPGDME < MDIPA (–121 ± 18) ~ DPGMME < DPG (–151 ± 12) ~ Glycerine ((–154 ± 12) at 383 K temperature. Note that the values of the adsorption energy are not a measure of the performance of a grinding aid (the correct measure is agglomeration energy) and also do not correlate with the volatility of the organic compounds (a possible measure is the cohesive energy of the pure liquid).

We computed agglomeration energies of dry and hydroxylated clinker surface ($C_3S$, $C_3A$) without and with GAs. Figures 2b and 2c give numbers of agglomeration energies of test and commercial GAs with chemical dosages of 0.20 mg/m$^2$. Experimental findings also follow trends on agglomeration energy.[1, 10, 12] We presented the experimental comparison of commercial grinding aids on the basis of normalized Blaine fineness and sieve residue at 32 μm with respect to data of MDIPA in the Table 1. Correlation of agglomeration energy with Blaine fineness and sieve residue of clinker particles is quite consistent to predict the performance of grinding aids. Grinding aids help to reduce surface forces between cleaved clinker particles which are very well captured by agglomeration studies.

These atomistic studies give an important insight into the molecular interactions; however the direct upscaling of results is only partially possible. For sure the mechanical, structural and surface properties of clinker phases can be embedded in an artificial clinker model that is used in Discrete Element Model simulations for clinker particle fragmentation. For the ball mill simulations with clinker particles, computed agglomeration energies can be important information.

**Table 1 Normalized Blaine fineness and sieve residue at 32 μm with respect to MDIPA values. Data were generated with the help of grinding experiments in a laboratory mill.**

| Compound | Dosage of GA [g/ton of cement] | Concentration of GA in water [wt. %] | Relative Blaine fineness [%] | Relative sieve residue at 32 μm [%] |
|---|---|---|---|---|
| MDIPA | 500 | 40 | 100.00 | 100.00 |
| TIPA | 500 | 40 | 95.80 | 107.80 |
| TEA | 500 | 40 | 92.70 | 112.10 |
| Glycerine | 500 | 40 | 90.50 | 118.50 |

Adsorption energy:    HC–TEA < HC–TIPA < HC–MDIPA < HC–Glycerine
Agglomeration energy: $C_3S$ > HC > HC–Glycerine > HC–TEA > HC–TIPA > HC–MDIPA
Grinding performance: Clinker < HC < HC–Glycerine < HC–TEA < HC–TIPA < HC–MDIPA

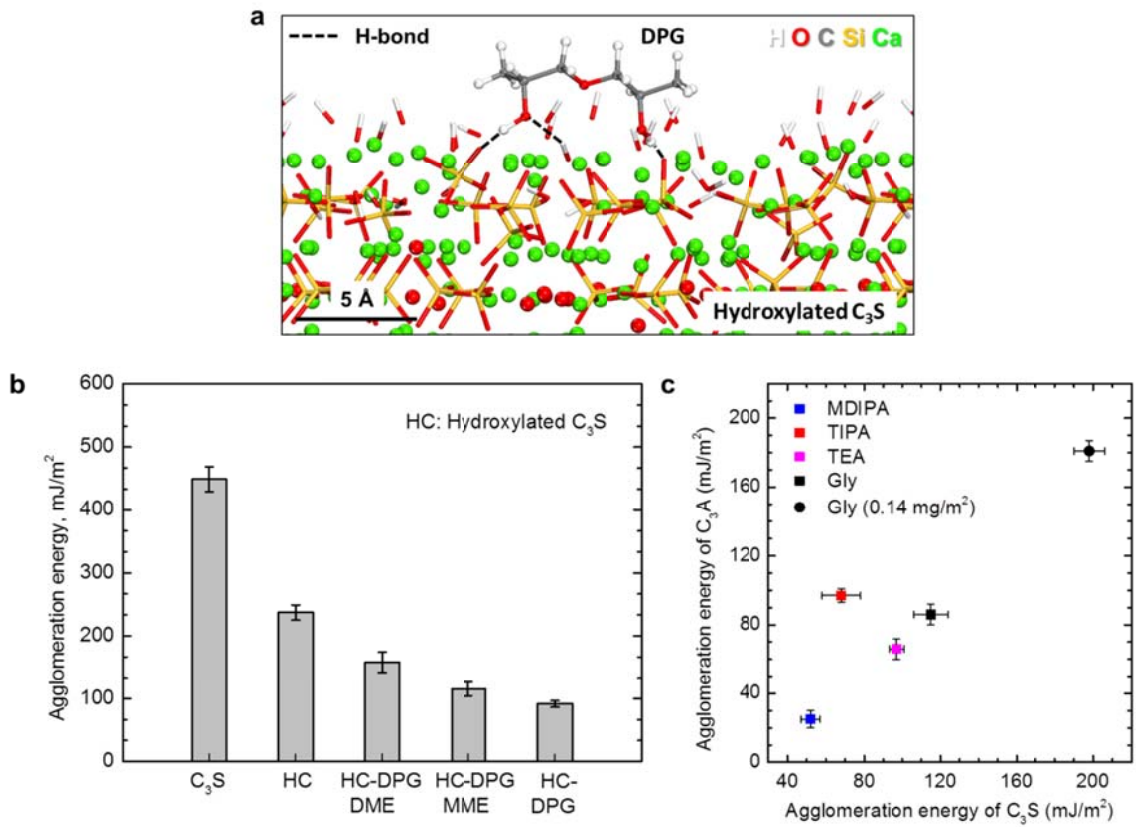

**2 (a) DPG molecule on the hydroxylated $C_3S$ surface with hydrogen bonds forming between hydroxyl groups of the molecules to silicate and hydroxide groups on the $C_3S$ surface at ball mill temperature 383 K. (b) Computed agglomeration energy of dry $C_3S$, hydrated $C_3S$, and in the presence of DPG, DPGMME, and DPGDME molecules on hydrated $C_3S$ surfaces at typical grinding temperature 363 K. (c) Reduced agglomeration energy due to the presence of commercial GAs with chemical dosage of 20 mg/m$^2$ for $C_3A$ and $C_3S$.[12] The effect of the additives is similar with a trend towards slightly higher agglomeration energy for $C_3S$.**

## Fragmentation of artificial clinker

We simulated composite breakable particles with crystalline and amorphous phases with respective material properties (Table A2) using the Discrete Element Model. Particle models with cohesive beam-truss elements are ideal for studying dynamic fracture and fragmentation problems. As various processes take place instantaneously during a short time span, the collaborative dynamics of particles, considering Hertzian collisional repulsive contacts with friction, cohesive interaction and volumetric forces, is solved by forward integration of the equation of motion of particles. Various fragmentation mechanisms, their origin, evolution and interaction during the process were studied in detail.[18-21] Clearly clinker is a multi-phase and multi-crystalline material, calling for a consideration of the heterogeneity of some sort. We simplify it by considering a single crystalline phase for $C_3S$ and $C_2S$ with elements organized in a hexagonal close packing, exhibiting intrinsic cleavage planes. As multiple crystallites are involved, first a packing of evolved single crystallites is constructed (Fig. 3a) from a collection of single crystals following the grain size distribution of $C_3S$ from a representative clinker microscopic section. Interstitial spaces are then filled by randomized particles to resemble an "amorphous phase".[22] Finally a spherical sample is cut from the system and impacted at different energy levels against a rigid target. In calibration simulations we match macroscopic elastic properties and fracture energies to the values obtained from force field calculations.

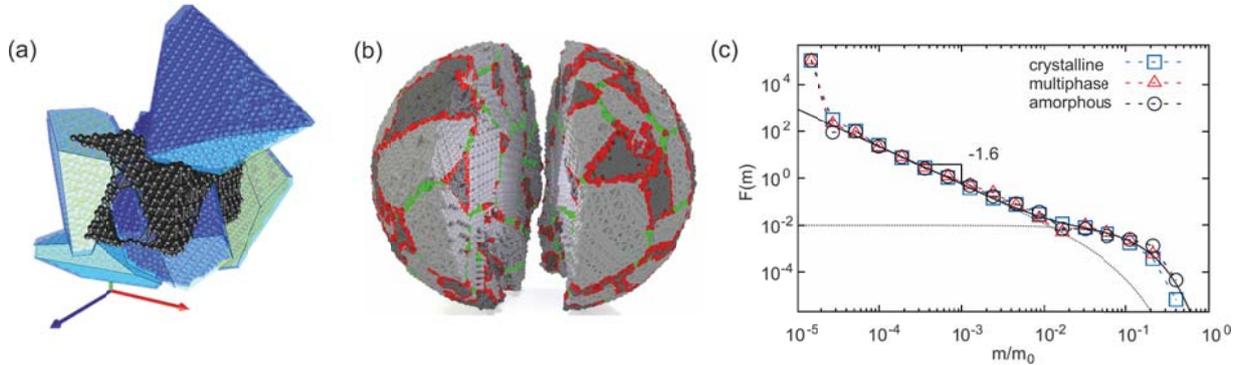

**3** (a) Crystalline phase for Alite ($C_3S$) and amorphous phase for aluminates ($C_3A$) and others. (b) Two largest fragments for multi-phase samples. Red bonds resemble amorphous-crystalline, green ones inter-crystalline interfaces. (c) Fragment mass distribution for different microstructures.[18, 22]

Beams are removed once their elliptical breaking rule of the von Mises criterion type:
$$(\varepsilon/\varepsilon_{th})^2 + \max(|\theta_i|,|\theta_j|)/\theta_{th} \geq 1 \qquad (1)$$
is fulfilled, where $\varepsilon$ is the longitudinal strain, $\theta_i$ and $\theta_j$ are the largest general end beam rotation angles connecting particle $i$ with $j$. Thresholds values ($\varepsilon_{th}$) are sampled from a Weibull distribution. The macroscopic strength of the material can be tuned by adopting the average breaking thresholds in each material phase separately so that the ratio between the ultimate strength of $C_3S$ and $C_3A$ follows the one of respective surface energies determined by molecular dynamics simulations. Additionally the breaking thresholds were set so that the ultimate strength of a micro sample $C_3A$ with dimensions in the range of the characteristic grain size is near 2 GPa.

Damage initiates inside spherical samples above the contact zone in a region where the circumferential stress field is tensile. Cracks initiated in this region grow to form meridional planes. If the collision energy exceeds a critical value which depends on the material's internal structure, cracks reach the sample surface resulting in fragmentation (see Figs. 3b and 4). We discovered that this primary fragmentation mechanism is surprisingly robust with respect to the internal structure of the material. Impact simulations were conducted for different collision energies. For the particular size of samples we simulated, we find that a critical collision energy of 0.8 mJ is necessary to fragment the system. Further increase of the impact energy produces more fragments with smaller sizes, but usually leaves at least one large fragment. For all configurations, a sharp transition from the damage to the fragmentation regime is observed, with critical collision energies being smallest for mono-crystalline samples. The mass distribution of the fragments follows a power law $P \propto m^{-\alpha}$ with an exponent $\alpha \approx 1.6$ for small fragments that is characteristic for the branching-merging process of unstable crack growth. Moreover this exponent depends only on the dimensionality of the system and not on the micro-structure (sees Fig. 3c). Hence attempts to reduce energy in comminution by manipulating the clinker micro-structure have a rather limited potential with respect to this fragmentation case.

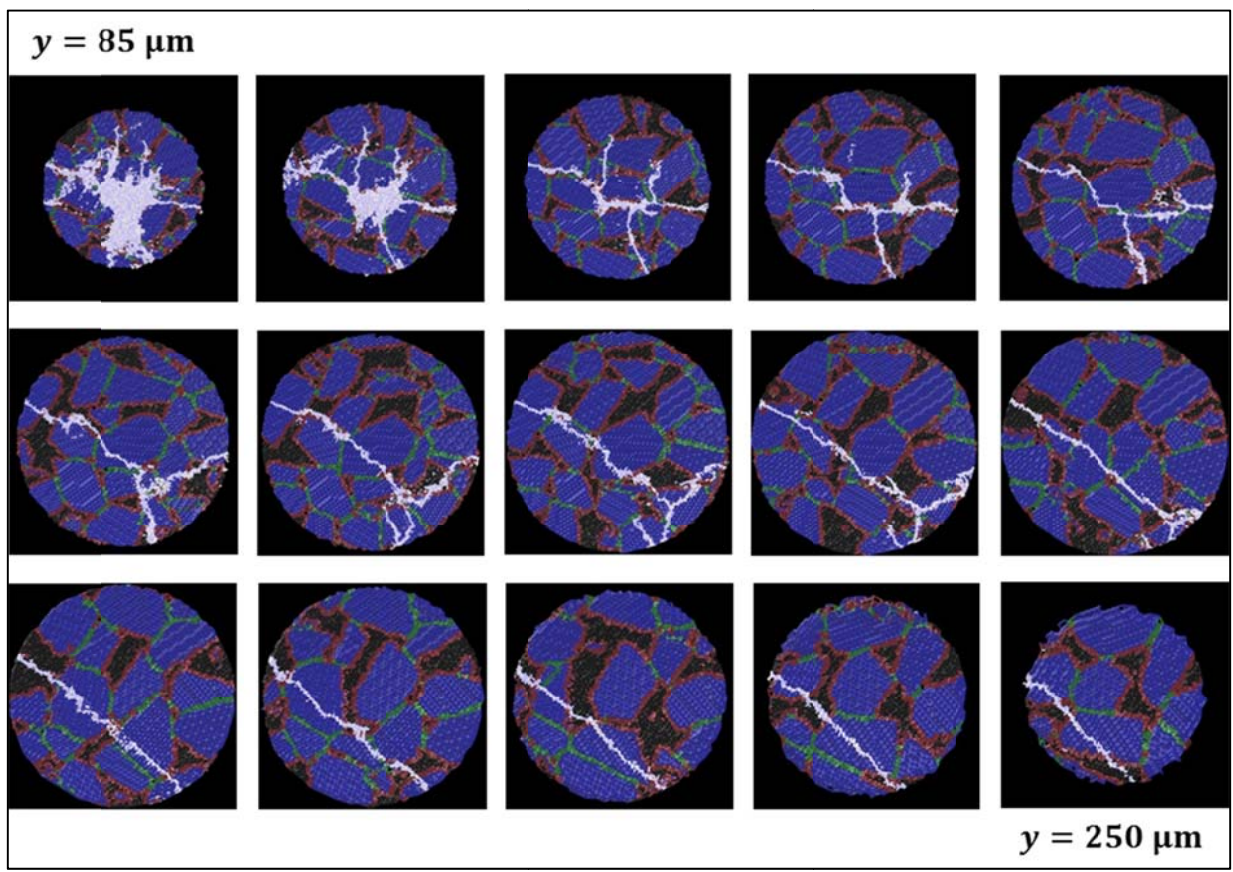

**4  Top view of slices of the sample at different y-positions, from $y = 85$ μm (top left) to $y = 250$ μm (bottom right). Bond colors correspond to material properties: black and red $C_3A$, blue and green $C_3S$, white are broken bonds.**

Energy dependent particle size distributions for impact fragmentation are important for the simulation of the outcome of ball mills, however one should note that fragmentation by impact above the critical energy is nothing but one possible way for size reduction of clinker as simulations of impact energy distributions reveal. Attrition, impact fatigue and fragmentation inside the granular bed with multiple contact points resemble cases that need to be studied separately to provide further realistic "fracture kernels". Also it needs to be noted that larger micro sections of clinker exhibit voids and cracks that point at large residual stresses that were not considered at this time.

## Simulation of ball mills

Simulations with Discrete Element Models have been employed by numerous authors to study particle dynamics in a variety of tumbling mills.[23-27] While a cement mill operator is principally interested in global behavior (e.g. particle size evolution during mill operation), this is determined by phenomena occurring at the microscopic scale, as described in earlier sections. To gain insights into how the presence of grinding aids influences the efficiency of the comminution process in an industrial cement mill, both a full-scale ball mill and laboratory mills have been considered. Although we are ultimately interested in predicting the behavior of a full-scale mill, laboratory mills are considerably more flexible and accessible for both experimental and numerical studies. Such mills are therefore used extensively to obtain detailed experimental measurements that can provide insights into the comminution process.[28, 29] In the proposed multi-scale numerical approach, laboratory mills are best used to study the evolution of the particle size distribution based on the previously described micro-scale breakage models. In the present study, particle breakage has not been modeled in the mill simulations. Of particular importance in this study, however, is the determination of the impact energy distribution, which not only provides a basis for comparison between mills of different sizes (i.e. the scaling of results)[30, 31], but also a means to link to the microscopic modelling of the breakage of individual clinker particles.

### Planetary mill

DEM simulations of the particle dynamics in both planetary and laboratory-scale ball mills have been carried out. A spring-dashpot model[32] has been employed to determine the normal and tangential collisional forces and torques acting on the particles. Presented here are results for a commercial planetary mill, the Fritsch Pulverisette 6, which is a small but convenient device for studying comminution. This device has a grinding bowl of 100 mm diameter rotating at 220 rpm and mounted on a disk counter-rotating at 400 rpm.[23] The ball charge consists of 25 steel balls of 20 mm diameter. The operation of a planetary mill can be modified by changing the rotational speed (while maintaining the same ratio of bowl and disk speeds), by changing the size and number of steel balls, and by modifying the clinker charge.

The purpose of DEM simulations was to analyze the effect of such operational changes. Fig. 5a shows an example of the instantaneous charge position and impact energy distributions computed for a planetary mill containing 100 g of clinker particles. The corresponding computed energy spectra are presented in Fig. 5b, where the dissipated energy is binned logarithmically with 20 bins per decade. Low-energy collisions between clinker particles and high-energy collisions involving ball-clinker contacts can be associated with the observed two peaks. This information can provide the impact energy required as input to the particle breakage modelling described in fragmentation study, enabling a prediction of the evolution of the particle size distribution due to comminution. Preliminary studies for the planetary mill indicated this approach to be promising, although its accuracy remains to be confirmed by validation studies.

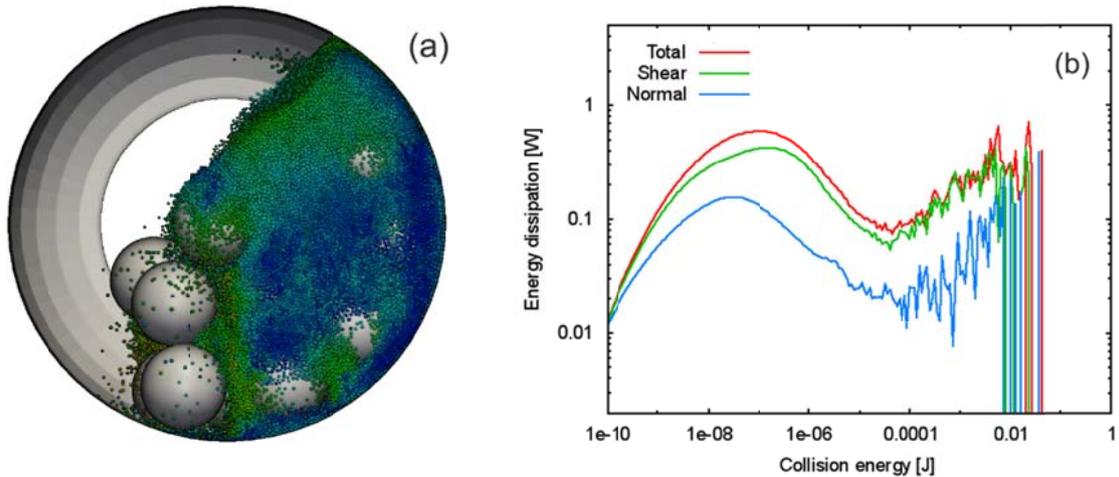

**5** Planetary mill: (a) ball charge, with clinker particles colored by their speed (blue: 0; red: 3 m/s), (b) impact energy distributions for normal, shear and total collisions.

**Cement mill**

An industrial cement mill generally consists of two chambers, the first containing large balls and a lifting liner, and the second containing smaller balls and a classifying liner. For the present study, a mill of 4.4 m diameter rotating at 15.2 rpm (~ 75% of critical speed) was considered. Although DEM simulations of the complete mill operation are not feasible with current computational resources, these requirements can be reduced by considering a short axial slice of each of the mill's chambers and including only the ball charge (i.e. neglecting clinker particles).[25, 26] These reduced DEM simulations can nevertheless provide representative impact energy distributions for a large-scale cement mill. Figure 6 presents a comparison of the IEDs computed for the two chambers of the cement plant mill and for a planetary laboratory mill. It can be noticed that the first chamber of the cement mill provides the higher collision energies necessary to break large clinker pieces, while the second chamber enables finer grinding at lower collision energies. It can also be observed from Fig. 6 that, by varying the rotational speed of the planetary mill, the peak of the collisional energy spectrum can be tailored to match that of the cement mill. Since the energy dissipated in clinker particle collisions is central to particle size reduction, it is reasonable to conjecture that comminution characteristics of the planetary mill will then be representative of those of the full-scale cement mill. The resulting evolution of the particle size distribution can then be predicted based on the results of breakage simulations proposed in particle fragmentation.

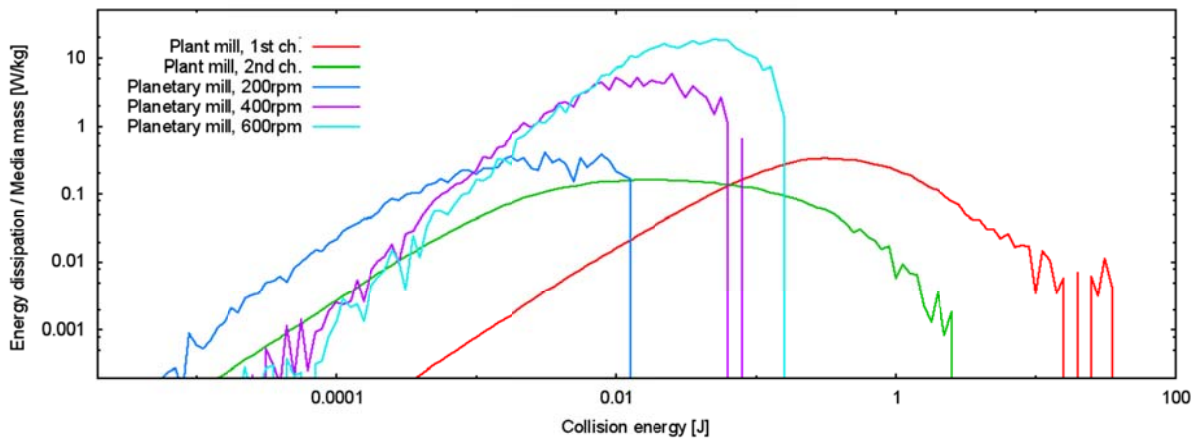

**6  Impact energy distributions computed for the two chambers of a cement plant ball mill and for a planetary mill for three different disk rotational speeds.**

## Summary and Outlook

We have combined diverse multi-scale modeling approaches to obtain new insights into the complex industrial problem of clinker comminution. Atomistic MD simulations quantitatively show that reducing the agglomeration energy is the main function of GAs. Micro-scale Discrete Element Model simulations of particle fragmentation with multi-phase properties obtained from the atomistic simulations give fragmentation thresholds and insight into the role of different phases. For impact fragmentation, crack planes are not substantially deflected at grain or phase boundaries, which is important with the reactivity of the cement in mind. Laboratory mills (especially the planetary mill) provide a convenient framework to couple the breakage dynamics of individual clinker particles with the global impact energy distribution computed using meso-scale DEM simulations. On the largest scale, the behavior of representative axial slices of the two chambers of a cement mill have been computed, proving that impact spectra can be comparable with those from laboratory mills. To summarize, the energy spectra obtained from DEM simulations provide a quantitative measure that forms the basis of scaling the behavior of a laboratory mill to that of an industrial mill.

A holistic numerical simulation methodology for industrial cement mills, including grinding media and fragmenting charge, will be most likely unattainable for a number of years. Nevertheless, today's computers are sufficiently powerful to identify and analyze important physical phenomena occurring at very different length scales. These phenomena can be studied in more detail by isolating the essential aspects from the global problem and studying them individually. As demonstrated in the present paper, atomistic simulations can not only play an important role in determining the relative efficiency of different GAs, but can also give quantitative information on properties of clinker phases and their surface interactions. For an ideal stress-free clinker this would suffice. However, in future works disorder and quenching of clinker should be simulated to relax the assumption of stress- and defect-free systems. Additionally studies of attrition, fatigue and bed fragmentation would enrich the fragmentation scenarios available for the more detailed simulation of comminution in cement mills.

## Acknowledgement

We acknowledge support by the Swiss Commission for Technological Innovation under grant no. KTI 13703.1 PFFLR-IW and by the Petroleum Research Fund of the American Chemical Society (54135-ND10).

## Appendix

**Table A1** Computed mechanical and surface properties of $C_3S$ and $C_3A$ phases of OPC clinker at temperature 298 K[10, 12].

| Mineral | Bulk modulus ($K$), GPa | Young's modulus ($E$), GPa | Poisson ratio |
|---|---|---|---|
| $C_3S$ | 105 ± 5 | 152 ± 5 ($E_x$), 176 ± 3 ($E_y$), 103 ± 11 ($E_z$) | 0.28 ± 0.08 |
| $C_3A$ | 98 ± 3 | 134 ± 5 | 0.29 ± 0.03 |
| | Cleavage energy, mJ/m$^2$ | | Temperature, K |
| $C_3S$ | 1340 ± 50 | | 298 |
| $C_3A$ | 1260 ± 50 | | 298 |
| | Agglomeration energy, mJ/m$^2$ | | Typical grinding temperature, K |
| Hyd. $C_3S$ | 240 ± 12 | | 363 |
| Hyd. $C_3A$ | 250 ± 30 | | 363 |

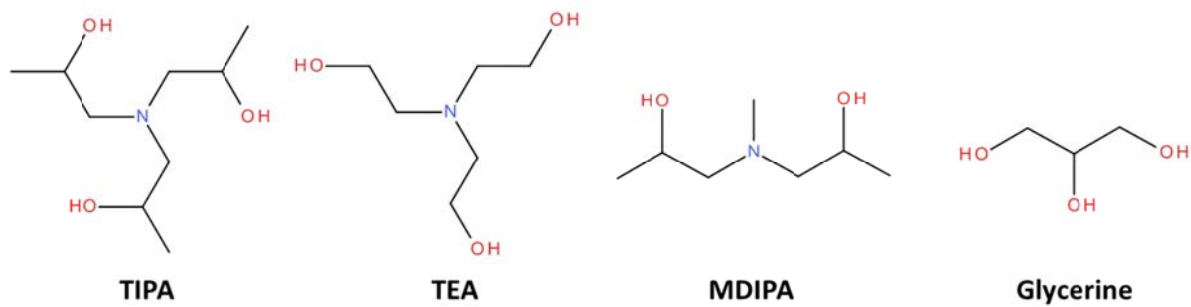

**Scheme A1** Chemical structure of commercial GAs used in cement industy such as triisopropanolamine (TIPA), triethanolamine (TEA), N-methyl-diisopropanolamine (MDIPA), and glycerine.

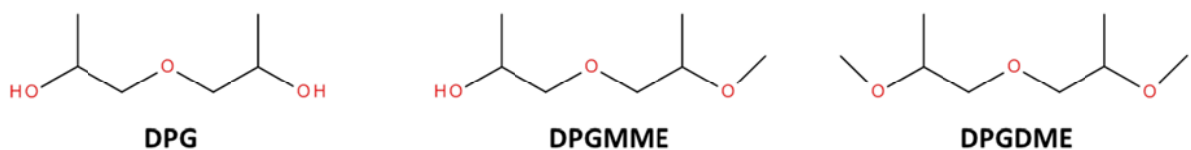

**Scheme A2** Chemical structure of organic additives which are also tested for their effectiveness during clinker grinding such as dipropylene glycol (DPG), dipropylene glycol monomethyl ether (DPGMME) and dipropylene glycol dimethyl ether (DPGDME).

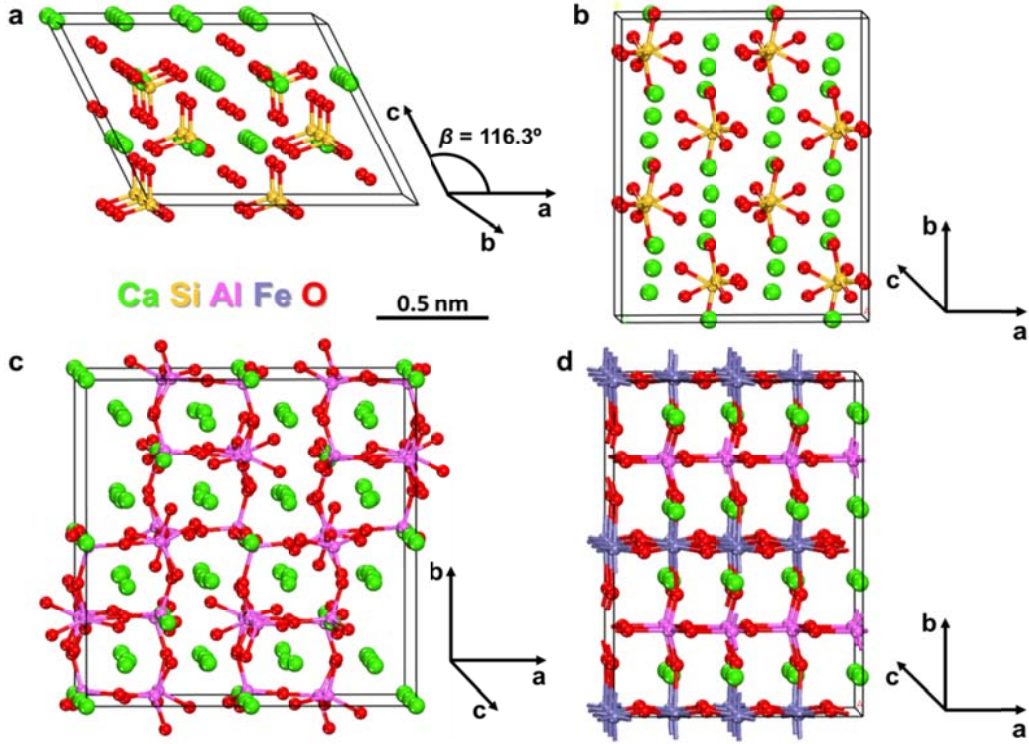

**Figure A1.** Crystal structure of clinker phases (a) M$_3$ polymorph of tricalcium silicate (1 × 2× 1 unit cells, a = 1.2235 nm).[33] (b) Monoclinic dicalcium silicate of the stable β polymorph (2 × 2 ×1 unit cells, a = 0.5502 nm).[34] (c) Unit cell of cubic lattice of tricalcium aluminate (1 × 1 × 1 unit cell, a = 1.5263 nm)[35] and (d) Super cell of tetracalcium aluminoferrite, (2 × 1 × 2 unit cells, a = 0.53369 nm).[36]

**Table A2 DEM model properties for the fragmentation of artificial clinker**

| Microscopic Property | | C$_3$S | C$_3$A | |
|---|---|---|---|---|
| Beams: | | | | |
| Stiffness | $E^b$ | 296 | 202 | $GPa$ |
| Average Length | $L_0$ | 5 | 5 | $\mu m$ |
| Diameter | $D^b$ | 3 | 3 | $\mu m$ |
| Strain Threshold | $\varepsilon_{th}$ | 0.025 | 0.02 | - |
| Bending Threshold | $\theta_{th}$ | 3.0 | 2.5 | ° |
| Shape parameter | $k$ | 10 | 10 | - |
| Elements: | | | | |
| Stiffness | $E^p$ | 296 | 202 | $GPa$ |
| Diameter | $D^p$ | 5 | 5 | $\mu m$ |
| Density | $\rho$ | 5000 | 5000 | $kg/m^3$ |
| Interaction: | | | | |
| Friction Coefficient | $\mu$ | 1 | | - |
| Damping Coefficient | $\gamma_n$ | $10^{-7}$ | | $Ns/m$ |
| Friction Coefficient | $\gamma_t$ | $10^{-8}$ | | $Ns/m$ |
| System: | | | | |
| Time increment | $\Delta t$ | $1 \times 10^{-10}$ | | $s$ |
| Number of elements | $N^p$ | 90 000 | - | |
| Number of beams | $N^b$ | 700 000 | - | |
| System Diameter | $D$ | 260 (spherical samples) | $\mu m$ | |
| Macroscopic Property | | | | |
| Stiffness | $E$ | $145 \pm 10$ | $135 \pm 10$ | $GPa$ |
| Poisson Ration | $\nu$ | 0.1 | 0.14 | - |
| Density | $\rho$ | 2700 | 3500 | $kg/m^3$ |
| Volume fraction | | 78 | 22 | % |
| Ultimate strength | $\sigma_{max}$ | 2.4 | 1.7 | $GPa$ |